# Correlative cellular ptychography with functionalized nanoparticles at the Fe L-edge

Marcus Gallagher-Jones[1†], Carlos Sato Baraldi-Dias[1†], Alan Pryor, Jr.[1†], Karim Bouchmella[2], Lingrong Zhao[1,3], Yuan Hung Lo[1], Mateus Borba Cardoso[2], David Shapiro[4], Jose Rodriguez[5*] and Jianwei Miao[1*]

[1]Department of Physics and Astronomy and California NanoSystems Institute, University of California Los Angeles, California 90095, USA. [2]Brazilian Synchrotron Laboratory (LNLS), Brazilian Center for in Energy and Materials, CEP 13083-970, Campinas, São Paulo, Brazil. [3]Department of Physics and Astronomy, Shanghai Jiao Tong University, Shanghai 200240, China. [4]Advanced Light Source, Lawrence Berkeley National Laboratory, Berkeley, California 94720, USA. [5]Department of Chemistry and Biochemistry, UCLA-DOE Institute for Genomics and Proteomics, University of California, Los Angeles, California, 90095, USA.

[†]These authors contributed equally to this work.
[*]Correspondence authors. jarod07@ucla.edu or miao@physics.ucla.edu.

## ABSTRACT

Precise localization of nanoparticles within a cell is crucial to the understanding of cell-particle interactions and has broad applications in nanomedicine. Here, we report a proof-of-principle experiment for imaging individual functionalized nanoparticles within a mammalian cell by correlative microscopy. Using a chemically-fixed, HeLa cell labeled with fluorescent core-shell nanoparticles as a model system, we implemented a graphene-oxide layer as a substrate to significantly reduce background scattering. We identified cellular features of interest by fluorescence microscopy, followed by scanning transmission X-ray tomography to localize the particles in 3D, and ptychographic coherent diffractive imaging of the fine features in the region at high resolution. By tuning the X-ray energy to the Fe L-edge, we demonstrated sensitive detection of nanoparticles composed of a 22 nm magnetic $Fe_3O_4$ core encased by a 25-nm-thick fluorescent silica ($SiO_2$) shell. These fluorescent core-shell nanoparticles act as landmarks and offer clarity in a cellular context. Our correlative microscopy results confirmed a subset of particles to be fully internalized, and high-contrast ptychographic images showed two oxidation states of individual nanoparticles with a resolution of ~16.5 nm. The ability to precisely localize individual fluorescent nanoparticles within mammalian cells will expand our understanding of the structure/function relationships for functionalized nanoparticles.

## INTRODUCTION

Functionalized nanoparticles are used in a broad array of nanomedicine applications for their utility as labels and drug delivery systems [1–4]. The ability to localize individual nanoparticles within cells is critical as it allows interactions between the nanoparticles and their target cells to be characterized, informing of the biological effects imparted by the nanoparticles. One method ideally suited to probe individual nanoparticles inside cells is coherent diffractive imaging (CDI) since it can image thick specimens with high resolution and contrast [5–23]. Since the first experimental demonstration in 1999 [5], various CDI methods have been developed [24] and a particularly powerful approach for imaging extended objects such as whole cells is ptychographic CDI (also known as ptychography) [25–41]. In ptychography, an extended sample is observed by illuminating with a coherent wave via a 2D raster scan. During such a 2D scan, diffraction patterns are recorded from overlapping fields of views with a pre-defined trajectory. The overlap between views can then be used as a strong constraint in phase retrieval algorithms [29], leading to a unique, robust reconstruction of the complex exit wave of the object and the illumination function [26,30–41]. Furthermore, by measuring diffraction intensity with a numerical aperture significantly higher than that of X-ray lenses, ptychography can reach spatial resolutions far beyond those of conventional X-ray microscopy. However, this powerful capability is hampered in the case of weakly scattering objects such as biological specimens, because the background scattering of the substrate and parasitic scattering from X-ray optics can dominate the weak signals from a biological specimen [30,36,39,41,42].

Here we demonstrate a correlative ptychographic approach for high-resolution imaging of functionalized nanoparticles internalized within an un-sectioned mammalian cell. To achieve high spatial resolution and image contrast, we first adapt x-ray transparent graphene-oxide substrates to support cells during the imaging process. These substrates offer a significant reduction in background scattering. Next, we label HeLa cells with core-shell nanoparticles, functionalized by the addition of a fluorescent moiety and a super-paramagnetic core, and identify a region of interest using fluorescence microscopy. Finally, we combine 3D localization of nanolabels within the cell, using scanning transmission X-ray microscopy (STXM) tomography, with correlative high-resolution ptychographic imaging, with enhanced contrast for the nanoparticles by tuning the X-ray energy to the Fe L-edge. This correlative cellular imaging method allows us to localize individual nanoparticles in a cellular context at multiple length-scales, ranging from tens of microns to the ten-nanometer level.

## RESULTS

### Nano-labeling of HeLa cells with fluorescent nanoparticles on a graphene-oxide substrate

HeLa cells were first grown on a substrate specifically engineered for high-contrast imaging, consisting of a commercially available gold TEM grid coated with lacey carbon (Ted Pella) on top of which we deposited a few layers of graphene-oxide (www.graphenesupermarket.com) using the drop casting method [43]. Graphene-oxide is biocompatible and allows for the adherent growth of HeLa cells while being effectively invisible to the soft X-ray probe (Fig. 1, Supplementary Fig. 1 and Materials and Methods). These grids provide significantly enhanced contrast when using the Nanosurveyor endstation in both STXM and ptychography modes (Supplementary Fig. 2). We attribute the improved stability and contrast to the atomically thin composition of the graphene layers and to the low density of graphene oxide relative to conventional silicon nitride-based supports. Graphene-oxide and silicon nitride have approximate densities of 1.9 g/cm$^3$ and 3.4 g/cm$^3$, respectively.

We imaged cells treated with fluorescent, core-shell mesoporous silica particles with an iron oxide core (Supplementary Fig. 3). Cells treated with nanoparticles were suspended in growth media for 30 minutes before gentle washing to remove excess nanoparticles, leaving mainly nanoparticles that had interacted with the cell surface or had been internalized. All samples were chemically fixed with paraformaldehyde, washed, and desiccated prior to imaging.

**STXM and Ptychography experiments**
X-ray experiments were performed on the Nanosurveyor instrument at BL 5.3.2.1 of the Advanced Light Source [44]. The X-ray energy was fixed at approximately 710 eV, the Fe L-edge. Incident X-rays were focused using a Fresnel zone plate with an outer diameter of 100 nm to give a total coherent flux of ~ $5 \times 10^5$ coherent photons $s^{-1}$ at the sample position. An order-sorting aperture was placed slightly upstream of the focal spot to remove all but the first order of the focused beam.

To facilitate a full range of rotation without obstructing the order sorting aperture, grids containing adhered cells and nano-labels were cut into strips thinner than 1mm using a zirconium nitride coated blade prior to transferring them to the Nanosurveyor instrument for X-ray experiments (Fig. 1). Thin sample strips allowed a greater range of accessible tilt angles for tomography experiments, which are normally restricted by the approximately 1 mm distance to the order-sorting aperture of the instrument (Fig. 1).

We acquired a total of 58 STXM projections with a 3 ms exposure and 50 nm step, giving a full field of view of 10×5 µm, with tilt angles ranging between -59° and +40° (Supplementary Fig. 4). Similarly, a total of 22 ptychography datasets, each consisting of 7,500 diffraction patterns, were collected using a compact Fast CCD [45]. Each ptychography pattern was measured with a dwell time of 200 ms and a step size of 60 nm, giving a full field of view of 9×3 µm with a pixel size of 5.5 nm (Supplementary Figs. 5 and 6).

**3D reconstruction of a HeLa cell's leading edge by STXM tomography**
STXM projections were aligned preliminarily by cross correlation. Background was subtracted from each projection by removing the average value in an empty region of the sample. The projections were normalized to have the same total sum, as the integrated 3D density of the sample should be constant. This tilt series, acquired from a section of an extended object, imposes a unique challenge for the tomographic reconstruction, as the sample is not isolated. The field of view can drift from one acquisition to the next, but the projection-slice theorem, a fundamental assumption of tomographic reconstruction techniques, assumes that the projected density in each image results from the same 3D volume. Therefore, isolating the same density in each projection presents a challenge and was overcome in the following way. The preliminary angular alignment, translational center, cropping, and intensity normalization parameters were used to make an initial set of projections.

The tomographic reconstruction was performed using a new generalized Fourier iterative reconstruction algorithm (GENFIRE)[46]. GENFIRE first pads zeros to each STXM projection and calculates an oversampled Fourier slice. The series of oversampled Fourier slices are interpolated to assemble a 3D Cartesian grid of the Fourier magnitudes and phases. The use of oversampling allows for accurate interpolation of grid points in the neighborhood of each Fourier slice[47], and the remaining Fourier grid points are set as undefined. The algorithm then iterates between reciprocal and real space using the fast Fourier transform (FFT) and its inversion. In reciprocal space, the measured grid points are enforced in each iteration, while undefined points are refined during the iterative process. In real space, the negative values and the electron density outside a pre-defined support are set to zero. An error metric, defined as the difference between the measured and calculated grid points, is used to monitor the convergence of the algorithm. The iterative process is then terminated when the error metric cannot be further reduced. From this preliminary 3D reconstruction, the alignment and cropping of each projection was optimized with another iterative refinement loop. For each experimentally acquired projection, the reconstructed 3D volume was back projected to a range of Euler angles about the current guess and aligned to the experimental projection by normalized cross correlation. The alignment with the maximum cross correlation yields updated values for the orientation and center of the projection. Next, a mask was made by smoothing and thresholding the best-fitting back projection, and from this mask a new input projection was obtained from the corresponding raw STXM projections. The purpose of this re-masking is to correct any inaccuracies in the initial cropping of the raw projections by utilizing the correlated information between the 3D reconstruction and each of its projections. After renormalization of the total intensity, an updated 3D reconstruction was computed from the new projections and orientation parameters. This loop was repeated until convergence of the alignment was obtained after 5 iterations. By using a Fourier based iterative process, GENFIRE produces better 3D reconstructions than other tomographic methods for a limited number of projections (Supplementary Fig. 7). A detailed comparison between GENFIRE and other 3D reconstruction methods will be presented in a follow-up paper.

**High-resolution ptychographic reconstruction**
To obtain the final ptychographic reconstructions, 7,500 raw diffraction patterns were first corrected for differences in offset and gain among the 192 CCD readout channels using software at BL 5.3.2.1. The corrected patterns were padded with zeros to give an image pixel size that, based on the scan step, would produce an integer number of pixels between scan positions to eliminate rounding errors. The diffraction patterns were then reduced in size by binning to 128x128 pixels. To remove bad frames caused by readout errors, the patterns were sorted by integrated intensity and frames with excessively high or low intensity removed. Finally, an intensity-based threshold was applied to the remaining patterns and used to provide an estimate of the incoherent background scattering. This average background was subtracted from the diffraction patterns and spuriously bright pixels were set to zero.

Initial ptychographic reconstructions were performed during the experiment using the SHARP algorithm to monitor data quality [48]. Reconstructions were later refined using a strip wise probe relaxation to account for artifacts introduced in some scans by the interaction of the beam with the order sorting aperture during data acquisition. This relaxation was implemented within a reconstruction scheme based on the extended ptychographic iterative engine (ePIE) [29]. The probe reconstructed via SHARP was used as an initial guess. The probe was updated at each scan position and monitored during the course of the reconstruction. After each macro cycle of ePIE, individual probes were averaged along the axis parallel to the tilt axis. Then a weighted average was performed between strips using a 1D Gaussian kernel to promote communication between different strips and to avoid striping artifacts (Supplementary Fig. 8). For these reconstructions, a strip width of 5 pixels and Gaussian kernel of 1.5 times the strip width gave the best results.

Phase normalization of the reconstructed images is necessary before comparison of neighboring ptychographic projections. A misalignment of the diffraction pattern relative to its true center translates to a phase gradient in the reconstruction of the object in real space [49]. Such a gradient can exist in both the horizontal and vertical directions, and an additional constant phase shift can exist without affecting the quality of fit of the reconstruction to the data. To perform this normalization, a pair of empty regions on either side of the sample was selected. A phase gradient in both the x- and y-axes was fit such that the standard deviation in the phase within these two regions was minimized. The constant phase term was then adjusted until the mean phase was zero within these regions.

**Correlative cellular imaging using fluorescent microscopy, STXM tomography and ptychography**

We achieved high-resolution imaging of the leading edge of a HeLa cell treated with fluorescent core-shell $Fe_3O_4$-$SiO_2$ nanoparticles by correlative microscopy. Fig. 2a shows part of an individual cell containing fluorescent nanoparticles. This same region was imaged using a coarse STXM scan to identify a smaller region of interest (Fig. 2b). A fine STXM scan was then performed on the region of interest and a tilt series of 58 STXM projections were acquired from this region. These projections were reconstructed to produce a 3D volume using GENFIRE, shown in Figs. 3, a, b and c. A small section of the leading edge of the HeLa cell was reconstructed in its entirety. Several regions within the reconstructed 3D volume show high absorption. We attribute this high absorption to the concentrated uptake of nanoparticles into the cell.

The resolution of our 3D reconstruction is sufficient to observe membrane ruffles near the upper leading edge of the cell (Fig. 3a). Due to obstruction of our sample by the TEM grid bars during rotation, our data suffers from a large missing wedge of more than 80°. Despite this, features in our reconstruction of the sample along the missing wedge direction are still well defined, allowing for precise 3D localization of the nanoparticles. Traditional tomographic reconstruction techniques, such as filtered back projection [50], suffer greatly from elongation artifacts when viewed along angles corresponding to unmeasured projections due to the anisotropic resolution of reciprocal space along this direction (Supplementary Fig. 7). The clarity of our results is attributed to our reconstruction algorithm, GENFIRE, which produces quality tomographic reconstruction from a limited number of measured projections. In this particular example, views along the missing wedge provide extended context that is critical for drawing biologically relevant conclusions regarding particle internalization.

To reveal the local distribution of fluorescent $Fe_3O_4$-$SiO_2$ nanolabels in or near the HeLa cell, we performed 2D ptychographic CDI on regions of interest identified from the 3D reconstruction using STXM tomography (Fig. 3, a, b and c). Our ptychographic scans produced high-resolution images that allow visualization of individual nanolabels in a correlative manner (Fig. 3, d, e and f). The images also show, with high contrast, fine features such as membrane ruffles, filopodia, and the thin lacey carbon support (Fig. 4). The enhanced contrast in these images is facilitated by the transparency of the graphene-oxide substrate and the presence of $Fe_3O_4$-$SiO_2$ nanolabels. The scattering signal produced by regions of cellular material containing these nanolabels is stronger than that produced by cellular material alone. The total scattered intensity in the presence of nanolabels increases by almost an order of magnitude compared to cellular material alone or the substrate (Supplementary Fig. 9). This increased image contrast and resolution facilitates localization of the nanolabels even when embedded within the cell.

Fig. 4 shows the phase and magnitude of the high-resolution ptychographic images. The phase images show fine cellular features such as membrane ruffles [51] and filopodia [52] (Fig. 4, a and b), while the magnitude images exhibit high contrast for the fluorescent $Fe_3O_4$-$SiO_2$ nanolabels. Further cellular detail, such as cytoskeletal components, are however obscured due to the limited resolution and poor contrast of sub-cellular structures in a background of other cellular components in 2D projection. This combination of the phase and magnitude images allows us to determine the accurate localization of the nanolabels near the cell periphery of, and inside the HeLa cell. By imaging close to an absorption edge, we also made possible the distinction of multiple oxidation states based upon recovered phase and absorption contrast of the nanoparticles. Our magnitude images allow the precise localization of the core-shell nanoparticles (Fig. 4, c and d). The phases of our ptychographic reconstructions show the presence of nanoparticle cores in two possible states, indicating potential differences due to oxidation (Fig. 4, b, Supplementary Fig. 10). This remains consistent across projections and is not an artifact of reconstruction (Supplementary Fig. 11). Particles in the $Fe^{3+}$ state, with an absorption resonance at 710 eV, will present zero phase shift relative to vacuum because of the value of the real part of the refractive index at the absorption resonance. Thus, those particles appear to have an annular structure in the phase images, which highlight the presence of the silica shell having non-zero phase shift. Although the initial step of the nanoparticle synthesis a single iron oxide is expected ($Fe_3O_4$). During the course of the storage at room temperature and in an aqueous medium, a subset of nanoparticles may naturally oxidize from $Fe_3O_4$ to $\gamma$-$Fe_2O_3$ [53]. The iron-oxide nanoparticles then exist as a mixture of these two oxides, each with distinct oxidative states and, consequently, distinct absorption coefficients at the X-ray energies used in this experiment. This difference in the scattering properties of our iron-oxide nanolabels is evident in the phase images (Fig. 4, b, Supplementary Fig. 10). While we did not focus on a spectroscopic analysis and did not image over a range of probe energies required to quantitatively measure the iron x-ray absorption coefficient of each species, our correlative technique is in principle capable of spectroscopic studies, allowing discernment of chemical species for a given nanoparticle in a cellular context.

**Resolution estimation**
To estimate the 3D resolution in STXM tomography, we compared independent reconstructions performed on two separate halves of the tilt-series. The Fourier shell correlation (FSC) between the two reconstructions is an approach commonly used in single-particle cryo-electron microscopy to estimate the 3D resolution [54]. Based on the criterion of FSC = 0.143, we determined the 3D resolution of the STXM tomography reconstruction to be 157 nm (Supplementary Fig. 12). Next, we estimated the resolution of our ptychographic reconstructions in two ways. First, we calculated the average phase retrieval transfer function (PRTF) for all of the patterns in a particular scan [55]. This gave us an upper and lower bound of the resolution as being between 25 and 15.5 nm in the full period based on a threshold of 0.5 (Supplementary Fig. 13). Second, we performed line scans across individual nanoparticles, which are well resolved and of known size (~73 nm with ~22 nm core). Line scans across the nanoparticle cores show that spacing of 3 pixels or greater can be easily distinguished, from which we estimate a resolution of ~16.5 nm in the full period (Supplementary Fig. 13). Taking both of these into account, we estimated the overall resolution in our images to be approximately 16.5 nm in the full period.

**Radiation dose and damage**
Radiation dose is fundamentally limiting to high resolution imaging experiments. Our current experiment does not benefit from the advantage of cryo-protection and therefore directly suffers from the imparted X-ray dose at 710 eV. We limited the total dose on the sample by first obtaining low-dose STXM images of the sample. Each STXM projection imparts an estimated dose of only 25 Gy, with a full tomogram imparting a dose of $1.45 \times 10^3$ Gy; this is well within the tolerable limit for our desired resolution in 3D [56]. A single ptychography scan imparted a dose of $1.17 \times 10^3$ Gy per projection on the sample, similar to the total dose for the STXM tomography dataset. For this reason, while a full ptychographic tomography series was captured, we limited our analysis to a subset of measured projections suspecting that the high dose could lead to changes in sample morphology during the course of the tomography series. The zero degree projections taken before and after the ptychographic tomography series confirm these changes. In contrast, the appearance and localization of nanoparticles, which are more tolerant to dose, remains unchanged (Supplementary Fig. 14). Since the total dose imparted to this cell is within the tolerable limit for cryogenically frozen samples [56], we believe that cryo-preservation alone will facilitate similarly high-resolution results in 3D. Furthermore, cryo-preservation will alleviate some of the issues related to interpreting cellular ultra-structure as the cells will be preserved in a more natural state. The combination of higher-resolution and more faithful preservation of cellular material will significantly benefit this imaging method.

**DISCUSSION**

The power of correlative microscopy lies in its potential to bridge the gap between complementary techniques, possibly expanding where a given method may be limited. Ptychographic CDI can provide high contrast images with a spatial resolution in the tens of nanometers from relatively thick, extended cells, particularly in the soft x-ray regime. As such, it is well suited to image in the regime between visible light and electron microscopy and its incorporation into correlative schemes is an area of active research [20,23,38,41]. However, biological specimens are a challenge to image with high resolution due to their sensitivity to radiation. This sensitivity imposes strict limits on 3D imaging of cells, given the high dose required to obtain multiple projections of a single cell. STXM tomography complements ptychographic imaging, providing an opportunity to image a radiation sensitive sample in 3D with a lower dose than a single ptychography scan. In doing so, STXM tomography provides an overview of a large sample region or an entire cell in 3D prior to high-resolution ptychographic imaging. In the ideal correlative experiment, a labeled sample could be initially inspected by fluorescence microscopy to identify temporal and spatial regions of interest. Then, a STXM tilt series would provide an overview of the whole cell in 3D. Next, high-resolution ptychographic imaging would provide near molecular details that could be correlated with known cellular structures. Finally, electron microscopy could be used to provide true molecular detail. The present revolution in imaging methods across the length scales will continue to benefit such correlative approaches.

In summary, we demonstrate a proof-of-principle correlative imaging method across multiple length scales of mammalian cells treated with functionalized fluorescent nanoparticles. Using a HeLa cell as a model system, we first identify cellular features of interest by fluorescent microscopy and correlate them in 3D via STXM tomography. We then image sub-regions of interest by ptychographic CDI with a resolution of ~16.5 nm. We observe fine biological features such as membrane ruffles and filopodia, and accurately localize individual fluorescent nanoparticles near the cell periphery and inside the cell. By choosing X-ray energies near the Fe L-edge, we enhance the image contrast of the core-shell $Fe_3O_4$-$SiO_2$ nanoparticles and identify them in two oxidation states. The ability to detect different oxidation states is important when the magnetic properties of the nanoparticles play a critical role in their function, such as magnetically induced heating of $Fe_3O_4$ nanoparticles in cancer therapy [3]. Although the nanoparticles were not targeted to specific biological structures in this study, incorporation of targeting moieties onto these nanoparticles is feasible and remains under investigation [57,58]. By incorporating cryogenic techniques, the applicability and resolution of this method can be further improved by removing the need for extensive fixation protocols [56]. Higher resolution would allow the visualization of sub-cellular organelles with much greater detail, such that the interaction between nanoparticles and organelles can be better understood. The ability to perform correlative cellular imaging and localize individual nanoparticles inside intact, un-sectioned mammalian cells through a combination of fluorescent microscopy, STXM tomography and ptychography will not only yield a more comprehensive understanding of the cell as a complex biological system, but also find applications in quantifying cell-particle interactions in nanomedicine.

## MATERIALS AND METHODS

**Synthesis and characterization of fluorescent $Fe_3O_4$-$SiO_2$ core-shell nanoparticles**
The Nanoparticles (NPs) synthesis protocol was adapted from the literature [59,60] to generate fluorescent $Fe_3O_4$-$SiO_2$ NPs as follows. A mixture of hydrated iron oxide (FeO(OH), 0.181 g), oleic acid (3.180 g) and docosane (5.016 g) was prepared in a round flask and stirred under vacuum for 1 h. This mixture was refluxed at 350 °C for 2 h in argon atmosphere. After cooling to room temperature, the resulting black solid was dissolved in pentane, mixed with an ether:ethanol solution (2:1) and centrifuged. The decomposed organic black solution was removed and the $Fe_3O_4$ nanoparticles were re-dispersed in pentane and washed with the ether:ethanol solution. After centrifugation, the NPs were stabilized with oleylamine (200 µL) and dispersed in chloroform. Then, 1.75 mL of 2 M NaOH solution was added into 200 mL of a CTAB solution (25 g.L$^{-1}$) and this mixture was kept at room temperature for 1 h under constant stirring. This mixture was heated up to 75 °C and 7.5 mL of $Fe_3O_4$ NPs chloroform suspension was slowly added into the CTAB basic aqueous solution under vigorous stirring. The formation of an oil-in-water microemulsion resulted in a turbid brown solution. The resulting solution was vigorously stirred for 2 h at 75 °C resulting in a transparent black $Fe_3O_4$/CTAB suspension. Then, 5 mL of tetraethylorthosilicate (TEOS) and 5 mL of FTIC-APTES (10 mg of fluorescein isothiocynanate [FITC] was reacted with 250 µL of 3- aminopropyltriethoxysilane [APTES] in 10 mL of ethanol for 2 h) were added to the $Fe_3O_4$/CTAB suspension and stirred for 3 h at 75 °C to obtain the fluorescent $Fe_3O_4$-$SiO_2$ NPs. These nanoparticles were washed 3 times with ethanol to remove the unreacted species and then dispersed in 20 mL of ethanol. The remaining sample was refluxed twice with an alcoholic solution of ammonium nitrate (6 g.L$^{-1}$, $NH_4NO_3$) at 50 °C to remove CTAB from the NPs pores. After washing them twice with ethanol, the fluorescent $Fe_3O_4$-$SiO_2$ NPs were dried under airflow for a few hours. All chemicals used along this protocol were purchased from Sigma-Aldrich. The morphology of the nanoparticles was investigated on a JEM 2100 JEOL transmission electron microscope. The samples were prepared by dropping a few drops of a highly diluted ethanolic nanoparticle suspension on regular TEM grids. The corresponding particle size distribution was obtained by measuring the core and the nanoparticle diameters of more than 400 particles.

**Preparation of graphene-oxide coated lacey carbon TEM grids**
All the biological assays were performed on a substrate composed of a Gold 400 Mesh TEM grid with lacey carbon. Additionally, a graphene layer was deposited on top of the lacey carbon to allow the culture of adherent cells on the TEM grid. The graphene layer was created by deposition of graphene oxide on the TEM grid by the 'drop-casting' method. A solution composed of 50% ethanol in double distilled water containing graphene oxide sheets is dropped on top of the TEM grid. After drying, TEM grids were annealed at 150 ºC for 30 min creating a homogeneous surface. This process also guarantees sterilization of the TEM grid. The prepared TEM grid was submerged in Dulbecco's modified Eagle's medium (DMEM) supplemented with 10% fetal calf serum and placed on the bottom of a cell culture dish. Next, suspended HeLa cells are dropped on the solution and maintained at 37 ºC in a 5% $CO_2$ atmosphere for 24 h. This procedure allows the cells to adhere to and grow on the graphene.

**Nanoparticle treatment**
The HeLa cells that had adhered to the graphene TEM grid were treated with the core-shell nanoparticles. The nanoparticles were suspended in distilled water (100µg/ml), and a few drops were added to the solution and dispersed gently. Cells were exposed to nanoparticles for 30 min, the medium was removed and the cells were gently washed with medium. Finally, the cells were fixed with paraformaldehyde.

**Acknowledgments:** We gratefully acknowledge the support of NVIDIA Corporation with the donation of the Quadro K5200 GPU used for this research. **Funding:** This work is supported by the National Science Foundation (DMR-1548924) and the DARPA PULSE program through a grant from AMRDEC. J.M. thanks the partial support by the Office of Basic Energy Sciences of the US Department of Energy (DE-SC0010378). C.S.B.-D. would like to acknowledge the partial support by Capes (CNPq 206124/2014-7). M.B.C. thanks the support of Fapesp (2013/22322-2) and K.B. acknowledges the support of Capes (CSF-PAJT - 88887.091023/2014-00). Soft X-ray ptychographic microscopy was carried out at beamline 5.3.2.1 at the Advanced Light Source, which is supported by the Director, Office of Science, and Office of Basic Energy Sciences, of the U.S. Department of Energy under Contract No. DE-AC02-05CH11231. This work is partially supported by the Center for Applied Mathematics for Energy Research Applications (CAMERA), which is a partnership between Basic Energy Sciences and Advanced Scientific Computing Research at the U.S Department of Energy. **Author contributions**: J.M. directed the research; J.M., M.B.C., J.A.R., and D.S. designed the experiment. M.G.-J., C.S.B.-D., A.P.Jr., K.B., L.Z., Y.H.L., D.S. and J.A.R conducted the experiments; M.G.-J., C.S.B.-D., A.P.Jr., J.A.R. and J.M. analyzed the data, performed reconstructions and interpreted the results. M.G.-J., C.S.B.-D., A.P.Jr., J.A.R. and J.M. wrote the manuscript with contributions from all authors. **Competing interests**: The authors declare that they have no competing interests. **Data and materials availability**: All data needed to evaluate the conclusions in the paper are present in the paper and the Supplementary Materials. Additional data related to this paper may be requested from the authors.

**Figures and Figure legends**

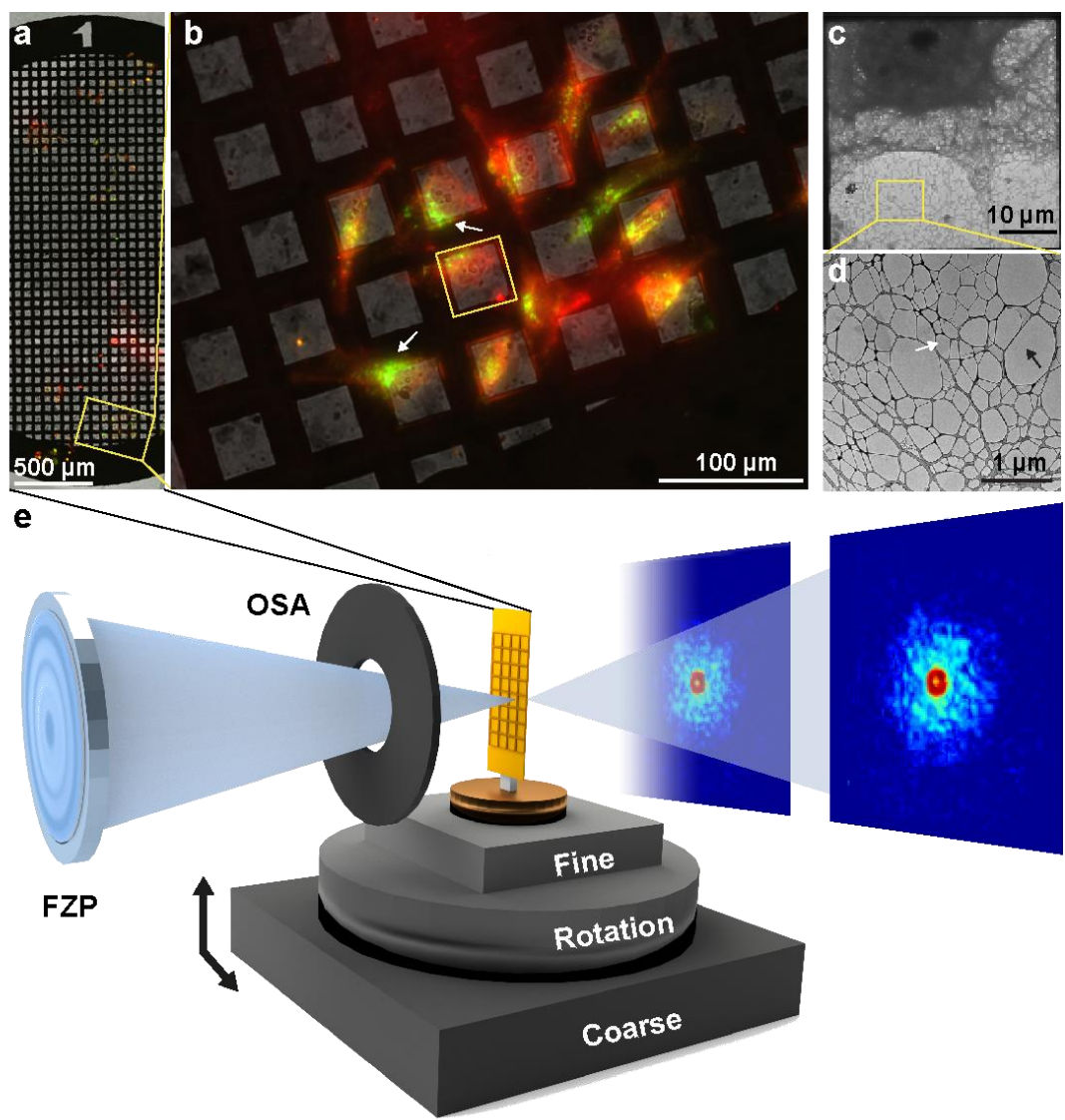

**Figure 1. Experimental setup for correlative microscopy**. (**a**) Composite fluorescent image of HeLa cells grown on graphene-oxide coated lacey carbon TEM grid. Cells were labeled with CMPTX (red) to facilitate tracking and treated with FITC labeled core-shell nanoparticles (green). (**b**) A magnified view of a region from this grid shows cells labeled with a tracking dye as well as fluorescent core-shell nanoparticles. White arrows point to cellular inclusions with clusters of fluorescent nanoparticles. (**c**) Electron micrograph of a portion of a HeLa cell covering an individual grid window, similar to the region highlighted in (**b**). (**d**) Magnified view of the lacey carbon grid. The black arrow points to empty regions of the grid whilst the white arrow indicates thin layers of graphene-oxide. (**e**) Experimental setup at BL 5.3.2.1 used for STXM/ptychographic imaging with key

components labeled. The X-ray beam is focused using a Fresnel zone-plate (FZP) and all but the first order focus blocked by an order-sorting aperture (OSA). The focused beam is rastered across the sample using high-precision stages under interferometric feedback and diffraction patterns are captured by a fast-CCD at each scan point.

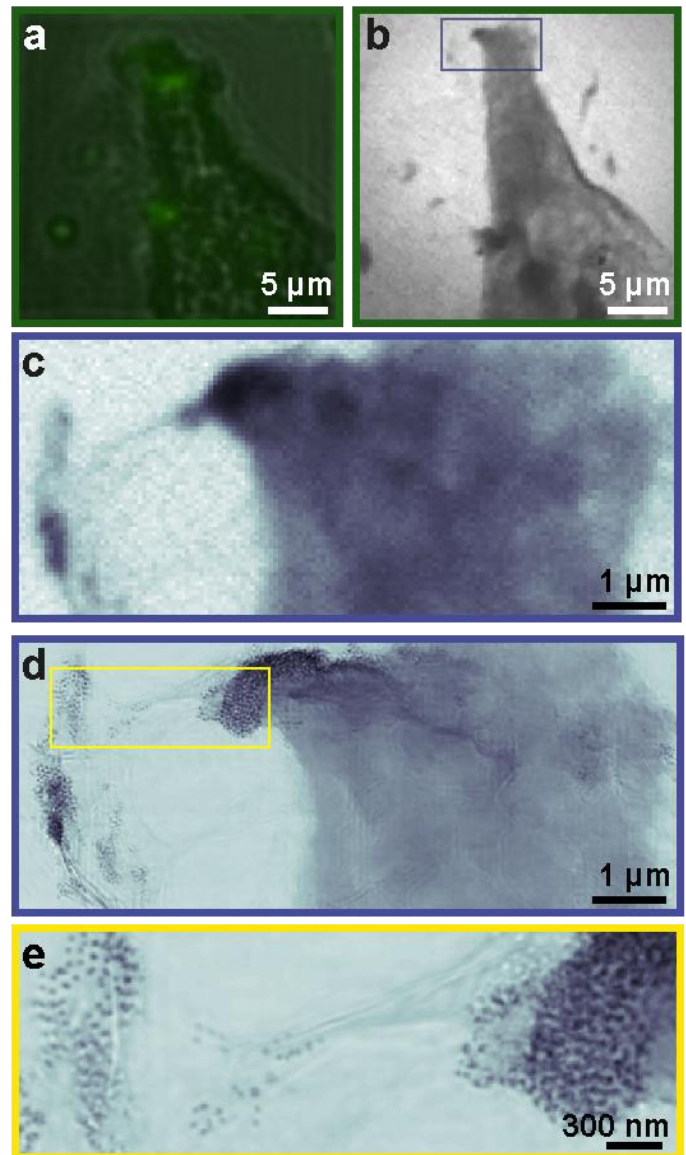

**Figure 2. Localization of functionalized nanoparticles in a cellular context with correlative microscopy.** (**a**) Part of a HeLa cell containing functionalized nanoparticles was first identified using fluorescent microscopy. (**b**) The same region imaged using a coarse STXM scan. (**c**) A fine STXM scan was then performed on a region of interest and a tomographic tilt series was acquired from this region. (**d**) Ptychographic imaging of the same region as (c) to obtain a higher resolution information. (**e**) Individual nanoparticles within and around the leading edge of the cell identified by the ptychographic reconstruction.

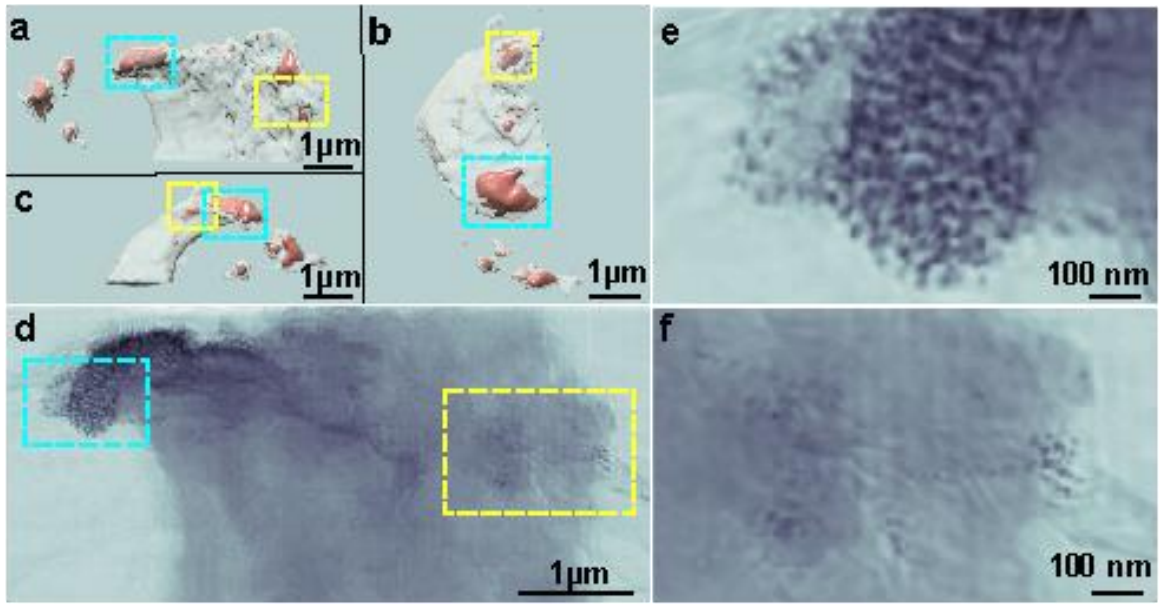

**Figure 3. STXM tomography reconstruction of the leading edge of a HeLa cell**. (**a-c**) Iso-renderings of the 3D reconstruction showing several high-density regions (orange) within the cell, viewed along the z-, minus y- and x-axes, respectively. (**d**) High-resolution ptychographic image confirming that the internalized high-density regions correspond to uptaken nanoparticles by the cell. (**e** and **f**) Magnified views of two regions in (**d**) labeled with cyan and yellow rectangles, respectively.

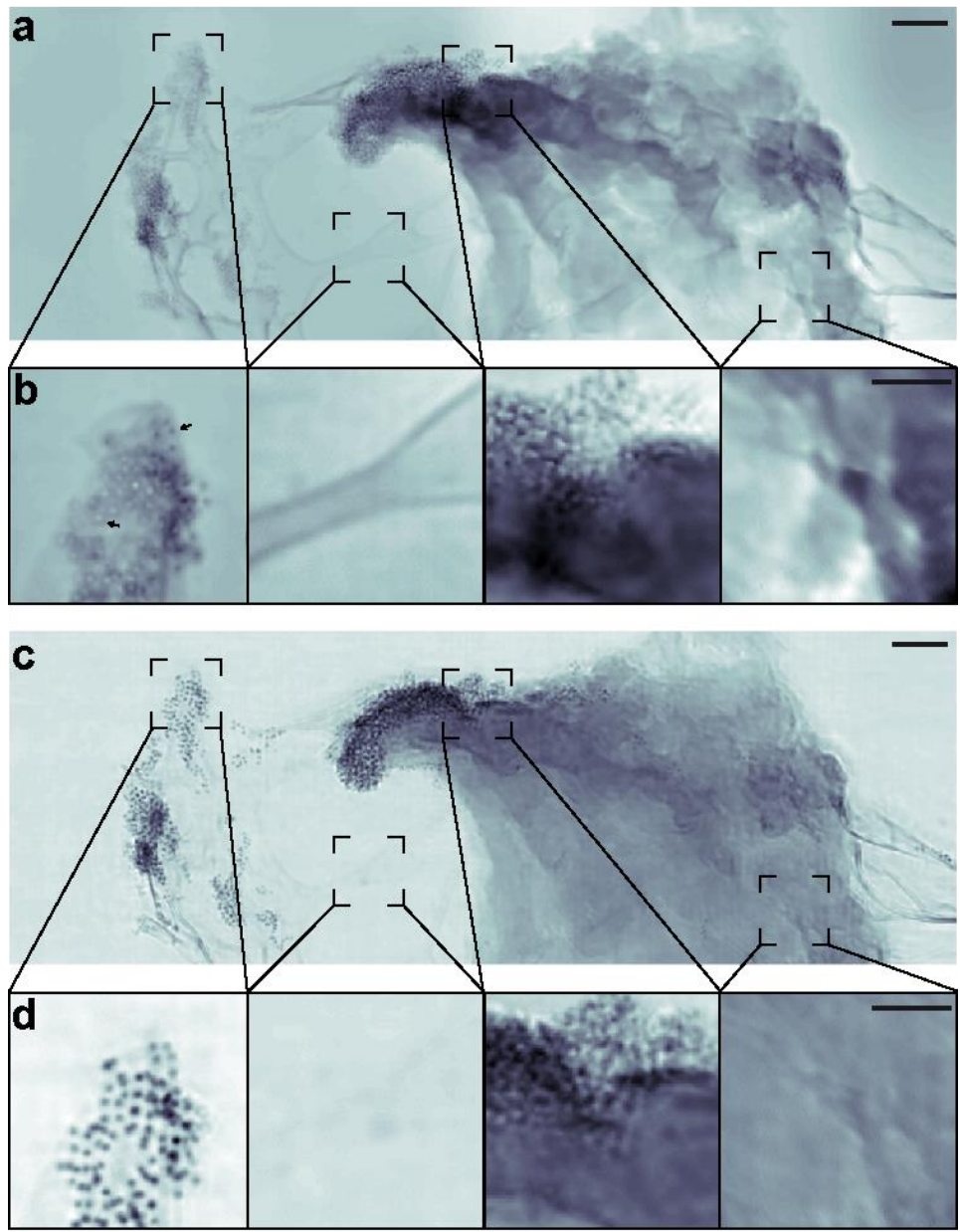

**Figure 4. Phase and magnitude ptychographic images of cellular structure with functionalized nanoparticles**. (**a**) Phase image of the ptychographic reconstruction of a HeLa cell labeled with core-shell nanoparticles showing high contrast for cellular features such as membrane ruffles and fillipodia. (**b**) Magnified views of the regions outlined by dashed boxes in (**a**), including (left to right) nanoparticles alone, graphene-oxide coated lacey carbon, cell and nanoparticles, and cell alone. The magnified view of the nanoparticles also demonstrates the phase's ability to discern the silica shell (light gray halo around cores indicated by black arrows) as well as the two different oxidation states (light and dark cores). A larger version of this can be seen in Supplementary Fig. 13. (**c**) Magnitude image of the ptychographic reconstruction showing high contrast for the $Fe_3O_4$ cores of the nanoparticles. (**d**) Magnified views of the same regions shown in (**b**), highlighting the different features that can be sharply resolved between the phase and magnitude images. Scale bars represent 500 nm (**a** and **c**) and 200 nm (**b** and **d**).

**Supplementary Materials**

Supplementary Fig. 1. Electron diffraction from Au-LaceyGO grids.

Supplementary Fig. 2. Comparison of STXM images on two different substrates.

Supplementary Fig. 3. Structural analysis of $Fe_3O_4$-$SiO_2$ (core-shell) nanoparticles.

Supplementary Fig. 4. STXM tomographic tilt series ranging from -59° to +40° in equal slope increments.

Supplementary Fig. 5. Magnitude images of the ptychographic tomography tilt series ranging from -59° to +40° in equal slope increments.

Supplementary Fig. 6. Phase images of the ptychographic tomography tilt series ranging from -59° to +40° in equal slope increments.

Supplementary Fig. 7. Comparison of tomographic reconstruction methods using STXM projections.

Supplementary Fig. 8. Removal of reconstruction artifacts by relaxation of probe uniformity.

Supplementary Fig. 9. Increase in scattering power in regions containing nanoparticles.

Supplementary Fig. 10. Magnified view of nanoparticles from the phase of the ptychographic reconstruction.

Supplementary Fig. 11. Consistency of nanoparticle phase contrast across projections.

Supplementary Fig. 12. Fourier shell correlation calculated from reconstructions of two half sets of the STXM tomography data.

Supplementary Fig. 13. Resolution estimates of ptychographic reconstructions

Supplementary Fig. 14. Comparison of the 0° projection before and after tomography series acquisition.